\documentclass[preprintnumbers,numbers,sort&compress,
               nofootinbib,3p, 
                            showpacs,
               colorlinks,
               linkcolor=blue,
               citecolor=blue]{elsarticle}

\usepackage{lineno,hyperref}
\modulolinenumbers[5]

\journal{Journal of \LaTeX\ Templates}
 \usepackage{graphicx}
\usepackage{dcolumn}
\usepackage{bm}
\usepackage{hyperref}

\usepackage{graphicx}
\usepackage{dcolumn}
\usepackage{bm}
\usepackage{hyperref}

\usepackage{amsmath}
\usepackage[caption=false]{subfig}
\usepackage{gensymb}

\def\ra{\rangle}
\def\la{\langle}
\def\rmd{\mathrm{d}}  

\newcommand{\be}{\begin{eqnarray}}
\newcommand{\ee}{\end{eqnarray}}
\newcommand{\beq}{\begin{equation}}
\newcommand{\eeq}{\end{equation}}
\newcommand{\exclude}[1]{}

\graphicspath{{./Figures/}} 

\begin{document}
    
    \begin{frontmatter}
    
       \title{Structure Formation Paradigm and    Axion Quark Nugget dark matter model }
       \author{  Ariel  Zhitnitsky}
       \address{Department of Physics and Astronomy, University of British Columbia, Vancouver, V6T 1Z1, BC, Canada}
     
      \begin{abstract}
           We advocate an idea that ``non-baryonic" dark matter   in form of nuggets   made of standard model  quarks and gluons (similar to the       old idea of the Witten's strangelets) could play a crucial role in structure formation. The corresponding macroscopically large objects, which are called the axion quark nuggets (AQN) behave as    {\it chameleons}: they  do not interact with the surrounding material in dilute environment, but they become strongly interacting objects in sufficiently dense environment. The AQN model was invented long ago with a single motivation to explain the observed  similarity     $\Omega_{\rm DM}\sim \Omega_{\rm visible}$ between  visible and DM components.  This relation represents a very  generic feature of this framework,  not sensitive  to any parameters of the construction.    We argue that the strong visible-DM interaction may dramatically modify the conventional structure formation pattern at small scales to contribute to a resolution of  a variety  of interconnected     problems (such as Core-Cusp problem, etc) which have been a matter of active research and debates in recent years. We also argue that the same visible-DM interaction at small scales  is always accompanied by a broad band diffuse radiation.  We speculate that the recently observed excesses of the   UV  emission  by JWST   at high redshifts and   by GALEX in our own galaxy  might be a direct manifestation of  this AQN-induced radiation.  We also speculate that the very same source of energy injection could contribute to the resolution of another long standing problem related to the Extragalactic Background Light (EBL) with known discrepancies in many frequency bands (from UV to optical, IR light and radio emissions). 
           
        \end{abstract}


\end{frontmatter}
	
\section{Introduction}\label{sec:introduction}

Observational precision  data gathered during the last quarter  of century  
have guided the development of the so called concordance cosmological 
model $\Lambda$CDM of a flat universe, $\Omega 
\simeq 1$, wherein the visible hadronic matter represents only $\Omega_B 
\simeq 0.05$ a tiny fraction of the total energy density, see recent review \cite{Turner:2022gvw},
and interesting historical comments \cite{Bertone:2016nfn}.
  Most of the 
matter component of the universe is thought to be stored in some unknown 
kind of cold dark matter, $\Omega_{DM} \simeq 0.25$. The largest  
contribution $\Omega_{\Lambda} \simeq 0.70$ to the total density  
  is  cosmological dark energy with negative pressure, another mystery which 
  will not be discussed here.
 
 There is a fundamental difference between dark matter
  and ordinary matter (aside from the trivial difference
 dark vs.  visible). Indeed, 
 DM played a crucial role in the formation of the present  structure in the universe.  Without dark matter, the universe would have remained too  uniform to form the galaxies.  
 Ordinary matter could not produce fluctuations to create any significant  structures   because it remains tightly coupled to radiation, preventing it from clustering, until  recent epochs.   
On the other hand, dark matter, which is not coupled to photons, would permit tiny  fluctuations  
 to grow for a long, long time  before the ordinary matter decoupled from radiation.  Then, the ordinary matter would  be rapidly drawn to the dense clumps of dark matter and form the observed structure.   
The required material is called 
the  Cold Dark Matter (CDM), and  the obvious candidates are weakly interacting fundamental 
particles of any sort which are long-lived, cold and collision-less, see old review article  \cite{Blumenthal:1984bp} advocating this picture, and more recent  review article \cite{Tulin:2017ara} with extended list of updates. 
While this model works very well on large scales, a number of discrepancies have arisen
between numerical simulations and observations on sub-galactic scales, see e.g. recent reviews 
 \cite{Tulin:2017ara,Salucci:2020eqo} and references on original papers therein.
Such discrepancies have stimulated numerous alternative proposals including, e.g. 
  Self-Interacting dark matter, Self-Annihilating dark matter,
  Decaying dark matter, and many others,
  see   \cite{Tulin:2017ara,Salucci:2020eqo}  and references therein. 
  There are  many other cosmological/astrophysical observations
   which  apparently  also  suggest  that  
 the standard assumption (that the  dark matter made of   
absolutely stable and  ``practically  non-interacting" fundamental particles)
is oversimplified. Some of the observations    that may be in conflict with the standard
 viewpoint are:
 
 $\bullet$ Core-Cusp Problem.  The disagreement of the observations with high resolution
   simulations is alleviated   with time, but some questions still remain 
     \cite{Tulin:2017ara,Salucci:2020eqo}.
 
  $\bullet$ Missing Satellite Problem. The number of dwarf galaxies in the Local group is smaller  than predicted by collision-less cold dark matter (CCDM) simulations. This  problem is also becoming  less dramatic  with time but some questions still remain
     \cite{Tulin:2017ara,Salucci:2020eqo}.
   
   $\bullet$ Too-Big-to-Fail Problem.  This  problem is also becoming  less dramatic  with time but some questions still remain
     \cite{Tulin:2017ara,Salucci:2020eqo}.

The problems  mentioned  above\footnote{\label{puzzles}There are many more similar problems and very puzzling observations. We refer to the review papers  \cite{Tulin:2017ara,Salucci:2020eqo} on this matter. There are also  different, but related  observations which  apparently  inconsistent with conventional picture of the structure formation, and which will be mentioned in section \ref{observations}.} occur as a result of comparison of the N-body simulations with observations.  Therefore, one could hope that these problems could be eventually resolved  when better and more precise simulations (for example accounting for the baryon feedback processes, such as  gas cooling, star formation, supernovae, active galactic nuclei) become available. However, there are some observations which are not based on comparison with N-body simulations, and which are very hard to understand within conventional CCDM paradigm. We list below some of such 
puzzling observations. This list is far from being complete, see footnote \ref{puzzles}.
   
   $\bullet$ DM-Visible Matter correlation shown on Fig.1 in ref.  \cite{Salucci:2020eqo} for the normal Spirals, dwarf Spirals, low surface brightness and the giant elliptical galaxies  is very hard to interpret unless DM interacts  with SM particles, see also  earlier works  \cite{Donato:2004af, Salucci:2007tm} where such relations had been originally discussed. 
   
    $\bullet$  Another manifestation  of the DM-Visible Matter correlation is presented on Fig.3 in ref. \cite{Salucci:2020eqo}
    which shows that the density kernel $K_c (r)$ defined as 
    \be
    \label{correlation}
    K_c (r)\equiv [\rho_{\rm DM}(r)\cdot \rho_{\rm visible} (r)] \sim {\rm const} ~~~~ {\rm  ~~~~~at} ~~~~~ r\simeq r_0
    \ee
    is almost a constant for all Spiral galaxies when computed at a specific point $r\simeq r_0$, which roughly coincides with observed   size of the core. In fact, $K_c (r_0)$ varies by only factor of 2 or so when  masses   vary by several orders of magnitude.
   The density itself may also vary by several orders of magnitude for different galaxies of different masses and sizes. Both these observations unambiguously suggest that the DM and visible matter components somehow know about each other, and start to interact strongly  at   small scales $r< r_0$, while DM behaves as conventional CCDM  for large scales $r> r_0$.

The main goal of the present studies is to argue that   the aforementioned discrepancies  (and many other related problems referred to  in footnote \ref{puzzles}.)  may be alleviated if dark matter is represented in form of the  composite,    nuclear density objects, the so-called Axion Quark Nuggets (AQN).  The   AQN DM  model  was suggested long ago in \cite{Zhitnitsky:2002qa}  with a single motivation to 
  explain the observed  similarity between the DM and   the visible matter densities in the Universe, i.e. $\Omega_{\rm DM}\sim \Omega_{\rm visible}$, which is very generic and model-independent consequence of the construction, see below.     This model is very  similar in spirit to the well-known Witten's quark nuggets \cite{Witten:1984rs} with several novel features which resolve   the previous fundamental problems of the construction \cite{Witten:1984rs}, to be discussed in next section. For now, in this Introduction we want to make  only two  important comments  on the AQN dynamics which are relevant for the present work:

1.    The AQNs    behave as chameleon-like   particles during the epoch of the structure formation with $z\in (5-15)$  when re-ionization epoch starts and ionization fraction 
  is expected to be large. This is because the AQN properties strongly depend on environment
  as we discuss   in section \ref{DM-coupling}. They have  all features of ordinary CCDM in the very dilute environment. However,  they become strongly interacting objects  when the ordinary visible matter density becomes  sufficiently  high, which is indeed the case in central regions of the galaxies.   The  interaction with    surrounding material becomes essential in this case.  Precisely this feature explains the observed correlation (\ref{correlation}) as  we shall argue below. The same visible-DM interaction   generates EM radiation in many frequency bands, including  the UV emission. One could   speculate that the  recent  JWST observations \cite{JWST,ALMA,Labbe,Boylan-Kolchin:2022kae}, which apparently detect some   excess of the UV radiation  from red-shifted  galaxies could be  a direct manifestation of this UV radiation. 

2. The very same interaction of the visible-DM components may lead to many observable effects   at present epoch at $z=0$ as well, including the excessive UV radiation.  In fact, the AQNs may be responsible for   explanation of the mysterious and puzzling observation  \cite{Henry_2014,Akshaya_2018,2019MNRAS.489.1120A} suggesting that there is a strong component of the diffuse far-ultraviolet (FUV) background    which is very hard  to explain by conventional physics in terms of  the dust-scattered starlight. Indeed,  
 the  analysis  carried out in \cite{Henry_2014,Akshaya_2018,2019MNRAS.489.1120A}     disproves   this conventional picture by demonstrating that the observed FUV radiation is highly symmetric being  independent of Galactic longitude in contrast with highly asymmetric localization of the brightest UV emitting stars.  It has been suggested in \cite{Zhitnitsky:2021wjb} that the puzzling  radiation could be originated from the AQN nuggets which indeed are uniformly distributed.  
 
 It is important  to emphasize that in this work with the main goal to   study   the AQN-induced effects which may impact the structure formation at $z\in (5-15)$ when ionization fraction $x_e$ is expected to be sufficiently large, 
 we use the same  basic parameters which had been previously used for variety of different applications in dramatically different environments, including  ref.  \cite{Zhitnitsky:2021wjb} with  explanation of  the observed puzzling FUV emission  at present time. In both cases (present time   and high redshifted epoch) the physics is the same and  is determined by  the coupling $[\rho_{\rm DM}(r)\cdot \rho_{\rm visible} (r)] $, which essentially enters  the observed correlation (\ref{correlation}).

Our presentation is organized as follows. In the next section \ref{AQN} we overview the basic  elements of the AQN construction with main focus on the key features relevant for the present studies.  Section \ref{DM-coupling} represents the main technical portion of this work where  we argue   that   AQNs may dramatically   modify  the domain in parametrical space when 
cooling is sufficiently efficient and galaxies may form. In section \ref{sect:correlation} we explain how the  observed correlation (\ref{correlation})
could emerge within the AQN scenario. In section \ref{sect:Heater} we   argue that  the AQN-induced processes always accompany  the UV radiation.   We  further speculate that the  recent  JWST observations \cite{JWST,ALMA,Labbe,Boylan-Kolchin:2022kae} could be  a direct manifestation of this UV radiation.   The  JWST observations at large $z$  are in fact very similar to mysterious FUV studies at present time  \cite{Henry_2014,Akshaya_2018,2019MNRAS.489.1120A} as reviewed   in \ref{sect:Heater1}. 
Finally, we conclude with  section \ref{conclusion} where we  list a number of other mysterious  observations 
 in  dramatically different environments (during BBN epoch, dark ages, and at present time: on the galactic, Solar  and  Earth scales) which could be explained within  the same framework with the same set of parameters. We also suggest many new tests, which 
are based on qualitative, model-independent consequences of our proposal, and  which can substantiate or refute this proposal.

 \section{The AQN   dark matter  model }\label{AQN}
 We overview  the fundamental  ideas of the AQN model in subsection \ref{basics},  while  in subsection 
 \ref{galaxy} we list    some  specific features of the AQNs  relevant for the present work.
 
 \subsection{The basics}\label{basics}
 The AQN construction in many respects is 
similar to the Witten's quark nuggets, see  \cite{Witten:1984rs,Farhi:1984qu,DeRujula:1984axn}. This type of DM  is ``cosmologically dark'' as a result of smallness of the parameter relevant for cosmology, which is the cross-section-to-mass ratio of the DM particles.
This numerically small ratio scales down many observable consequences of an otherwise strongly-interacting DM candidate in form of the AQN nuggets.  

There are several additional elements in the AQN model in comparison with the older well-known and well-studied  {theoretical} constructions \cite{Witten:1984rs,Farhi:1984qu,DeRujula:1984axn}. First, there is an additional stabilization factor for the nuggets provided by the axion domain walls which  are copiously produced  during the   QCD  transition. This additional element    helps to alleviate a number of  problems with the original Witten's  model{\footnote{\label{first-order}In particular, a first-order phase transition is not a required feature for nugget formation as the axion domain wall (with internal QCD substructure)  plays the role of the squeezer. Another problem of the old construction  \cite{Witten:1984rs,Farhi:1984qu,DeRujula:1984axn} is that nuggets likely evaporate on the Hubble time-scale. For the AQN model, this is not the case because the vacuum-ground-state energies inside (the color-superconducting phase) and outside the nugget (the hadronic phase) are drastically different. Therefore, these two systems can coexist only in the presence of an external pressure, provided by the axion domain wall. This should  be contrasted with the original model \cite{Witten:1984rs,Farhi:1984qu,DeRujula:1984axn}, which is assumed to be  stable  at zero external pressure. 
This difference has dramatic observational consequence- the Witten's nugget will turn a neutron star (NS) into the quark star if it hits the NS. In contrast,  a  matter type AQN   will not turn an entire star into a new quark phase because the  quark matter in the AQNs   is supported  by external axion domain wall pressure, and therefore, can be extended only to relatively small distance $\sim m_a^{-1}$,   which is much shorter  than the NS size.  }.  
Secondly,  the nuggets can be made of {\it matter} as well as {\it antimatter} during the QCD transition. 

The presence of the antimatter nuggets in the AQN  framework is an inevitable and the direct consequence of the 
    $\cal{CP}$ violating  axion field  which is present in the system during the  QCD time. As a result of this feature      the DM density, 
    $\Omega_{\rm DM}$, and the visible    density, $\Omega_{\rm visible}$, will automatically assume the  same order of magnitude densities  
    $\Omega_{\rm DM}\sim \Omega_{\rm visible}$  irrespectively to the parameters of the model, such as the axion mass $m_a$. 
 This feature represents a generic property of the construction   \cite{Zhitnitsky:2002qa} as both component, the visible, and the dark are proportional to one and the same fundamental dimensional constant of the theory, the $\Lambda_{\rm QCD}$. 
  
  We refer to the original papers   \cite{Liang:2016tqc,Ge:2017ttc,Ge:2017idw,Ge:2019voa} devoted to the specific questions  related to the nugget's formation, generation of the baryon asymmetry, and  survival   pattern of the nuggets during the evolution in  early Universe with its unfriendly environment. We also refer to a recent brief review article \cite{Zhitnitsky:2021iwg} which explains a number of subtle points on the formation mechanism,  survival pattern of the AQNs during the early stages of the evolution,  including the Cosmic Microwave Background (CMB)  Big Bang Nucleosynthesis (BBN), and recombination epochs. 

 \begin{table*}
\captionsetup{justification=raggedright}
	\begin{tabular}{cccrcc} 
		\hline\hline
		  Property  && \begin{tabular} {@{}c@{}}{ Typical value or feature}~~~~~\end{tabular} \\\hline
		  AQN's mass~  $[M_N]$ &&         $M_N\approx 16\,g\,(B/10^{25})$     \cite{Zhitnitsky:2021iwg}     \\
		   baryon charge constraints~   $ [B]  $   &&        $ B \geq 3\cdot 10^{24}  $     \cite{Zhitnitsky:2021iwg}    \\
		   annihilation cross section~  $[\sigma]$ &&     $\sigma\approx\kappa\pi R^2\simeq 1.5\cdot 10^{-9} {\rm cm^2} \cdot  \kappa (R/2.2\cdot 10^{-5}\rm cm)^2$  ~~~~     \\
		  density of AQNs~ $[n_{\rm AQN}]$         &&          $n_{\rm AQN} \sim 0.3\cdot 10^{-25} {\rm cm^{-3}} (10^{25}/B) $   \cite{Zhitnitsky:2021iwg} \\
		  survival pattern during BBN &&       $\Delta B/B\ll 1$  \cite{Zhitnitsky:2006vt,Flambaum:2018ohm,SinghSidhu:2020cxw,Santillan:2020lbj} \\
		  survival pattern during CMB &&           $\Delta B/B\ll 1$ \cite{Zhitnitsky:2006vt,Lawson:2018qkc,SinghSidhu:2020cxw} \\
		  survival pattern during post-recombination &&   $\Delta B/B\ll 1$ \cite{Ge:2019voa} \\\hline
	\end{tabular}
	\caption{Basic  properties of the AQNs adopted from \cite{Budker:2020mqk}.} 
	\label{tab:basics}
\end{table*}

We conclude this brief review subsection with Table\,\ref{tab:basics} which summarizes the basic features and parameters of the AQNs.
 The parameter $\kappa$ in Table\,\ref{tab:basics}   is introduced to account for the fact that not all matter striking the nugget will 
annihilate and not all of the energy released by annihilation will be thermalized in the nuggets. The ratio $\Delta B/B\ll 1$ in the Table implies that only a small portion $\Delta B$  of the total (anti)baryon charge $  B$  hidden in form of the AQNs get annihilated during big-bang nucleosynthesis (BBN), Cosmic Microwave Background (CMB), or post-recombination epochs (including the galaxy and star formation), while the dominant portion of the baryon charge survives until the present time.  
 Independent analysis \cite{Santillan:2020lbj} and  \cite{SinghSidhu:2020cxw}    also support our original claims as cited in the Table\,\ref{tab:basics} that the anti-quark nuggets survive the BBN and CMB epochs. 
 
 Finally, one should mention here that the AQN model with the same set of parameters may explain a number of other puzzling observations 
in  dramatically different environments (during BBN epoch, dark ages, and at present time: on the galactic, Solar  and  Earth scales) as highlighted  in concluding section \ref{conclusion}.

 \subsection{When the AQNs start to interact in the galactic environment}\label{galaxy}
    For our present work, however,  the  most relevant studies  are related to the effects which  may occur when the AQNs made of antimatter 
    propagate in the environment with sufficiently large visible matter density $\rho_{\rm visible} (r)$ entering (\ref{correlation}). 
   In this case  the annihilation processes start and 
      a large amount of energy   will be injected to surrounding material, which may  be manifested in many different ways. What is more important for the present studies is that the same annihilation processes will dramatically reduce the ionization portion of the material $x_e$ during the galaxy formation at a redshift $z\in (5-15)$ because the ions are much more likely to interact  with the AQNs in comparison with neutral atoms due to the long-ranged Coulomb attraction.  
      
      The related  computations on the AQN-visible matter interaction originally have been carried out in \cite{Forbes:2008uf}
 in application to the galactic neutral environment at present time  with a typical density of surrounding   baryons of order $n_{\rm galaxy}\sim   {\rm cm^{-3}}$ in the galaxy, similar to the density to be discussed in the present work  at a redshift $z\in (5-15)$. We review  the computations \cite{Forbes:2008uf} with few additional elements which must be implemented in case of propagation of the AQN when galaxies just starting  to form.  

 We draw the AQN-structure on Fig \ref{AQN-structure}, where we use typical parameters from the  Table\,\ref{tab:basics}. There are several   distinct length scales of the problem: $R\sim 10^{-5}$ cm represents the size of the nugget filled by quark matter with $B\sim 10^{25}$ in CS phase. Much larger scale  $R_{\rm DW}\sim m_a^{-1}$  describes the axion DW   surrounding the quark matter. The axion DW has the QCD substructure surrounding the quark matter and  which has typical width of order $R_{\rm QCD}\sim 10^{-13} \rm cm$. Finally, there is always electro-sphere which represents a  very generic feature of quark nuggets, including the Witten's original construction. In case of antimatter-nuggets the electro-sphere comprises the positrons.  The  typical size of the electrosphere is order of $10^{-8} \rm cm$, see below. 
       \begin{figure}[h]
	\centering
	\captionsetup{justification=raggedright}
	\includegraphics[width=0.8\linewidth]{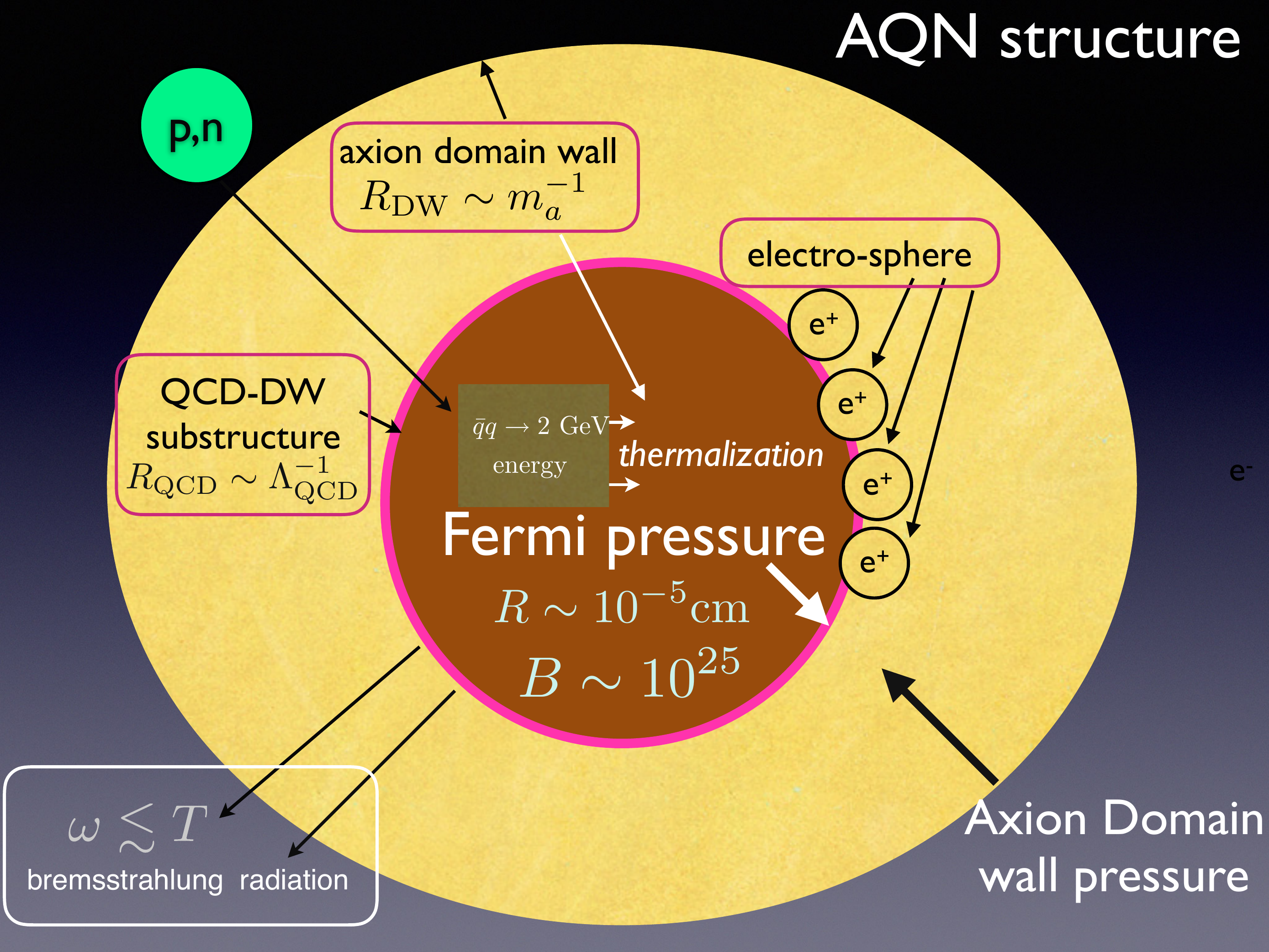}
		\caption{AQN-structure (not in scale), adopted from \cite{Zhitnitsky:2022swb}. The dominant portion of the energy $\sim 2$ GeV produced as a result of  a single  annihilation process inside the anti-nugget is released in form of the bremsstrahlung radiation with frequencies $\omega\leq T$, see description and notations in the main text.}
\label{AQN-structure}
\end{figure}

When the AQN enters the region of the baryon density $n $ the annihilation processes start and the internal temperature increases. 
  A typical internal temperature  $T$ of  the  AQNs for very dilute galactic environment can be estimated from the condition that
 the radiative output   must balance the flux of energy onto the 
nugget 
\be
\label{eq:rad_balance}
    F_{\rm{tot}} (T) (4\pi R^2)
\approx \kappa\cdot  (\pi R^2) \cdot (2~ {\rm GeV})\cdot n \cdot v_{\rm AQN},  
\ee 
where $n$ represents the baryon number density of the surrounding material, and $F_{\rm{tot}}(T) $ is total  surface emissivity, see below. 
The left hand side accounts for the total energy radiation from the  AQN's surface per unit time   while  
 the right hand side  accounts for the rate of annihilation events when each successful annihilation event of a single baryon charge produces $\sim 2m_pc^2\approx 2~{\rm GeV}$ energy. In Eq.\,(\ref{eq:rad_balance}) we assume that  the nugget is characterized by the geometrical cross section $\pi R^2$ when it propagates 
in environment with local baryon density $n$ with velocity $v_{\rm AQN}\sim 10^{-3}c$.
The factor $\kappa$  accounts    for large theoretical uncertainties related to the annihilation processes 
of the (antimatter)  AQN  colliding with surrounding material.

        The total  surface emissivity due to the bremsstrahlung radiation from electrosphere at temperature   $T$ has been computed in \cite{Forbes:2008uf} and it is given by 
\begin{equation}
  \label{eq:P_t}
  F_{\rm{tot}} \approx 
  \frac{16}{3}
  \frac{T^4\alpha^{5/2}}{\pi}\sqrt[4]{\frac{T}{m}}\,,
\end{equation}
 where $\alpha\approx1/137$ is the fine structure constant, $m=511{\rm\,keV}$ is the mass of electron, and $T$ is the internal temperature of the AQN.  
 One should emphasize that the emission from the electrosphere is not thermal, and the spectrum is dramatically different from blackbody radiation.
    From (\ref{eq:rad_balance})  one can estimate a typical internal nugget's temperature  when density $n$ assumes the typical values $n\sim \rm cm^{-3}$  relevant for this work:
 \be
 \label{T}
 T\sim 0.4 ~{\rm eV} \cdot \left(\frac{n }{\rm cm^{-3}}\right)^{\frac{4}{17}}\cdot\left(\frac{v_{\rm AQN}}{10^{-3}c}\right)^{\frac{4}{17}}  \cdot \kappa^{\frac{4}{17}}.
 \ee 
It strongly depends    on unknown parameter $\kappa$ as mentioned above. 
In case which is relevant for our studies when the    surrounding material   is a highly ionized  plasma   the parameter $\kappa$ effectively gets much larger  as the AQN (being negatively charged, see below)  attracts more positively charged ions from surrounding material. This attraction    consequently   effectively increases the cross section and the rate of annihilation,  eventually resulting in a larger value of $T$.  

Another  feature   which is relevant for our  present studies is the ionization  properties of the AQN. Ionization, as usual,   occurs in a system   as a result of  the high internal temperature $T$. In our case of the AQN characterized by temperature (\ref{T})     a large number of weakly bound positrons $\sim Q$ from the electrosphere   get excited and can easily leave the system. The corresponding parameter $Q$ can be estimated as follows: 
\be
  \label{Q}
Q\approx 4\pi R^2 \int^{\infty}_{0}  n(z, T)\rmd z\sim \frac{4\pi R^2}{\sqrt{2\pi\alpha}}  \left(m T\right)   \left(\frac{T}{m}\right)^{\frac{1}{4}} , ~~
  \ee
 where $n(z, T)$ is the local density of positrons at distance $z$ from the nugget's surface, which has been computed in the mean field approximation  in \cite{Forbes:2008uf} and has the following form
\begin{equation}
\label{eq:nz0}
n(z, T)=\frac{T}{2\pi\alpha}\frac{1}{(z+\bar{z})^2}, ~~ ~~\bar{z}^{-1}\approx \sqrt{2\pi\alpha}\cdot m  \cdot \left(\frac{T}{m}\right)^{\frac{1}{4}}, ~~~~  \end{equation}
where $\bar{z}$ is the integration constant is chosen to match the Boltzmann regime at sufficiently  large $z\gg \bar{z}$. Numerical 
studies ~ \cite{Forbes:2009wg}  support the approximate analytical  expression (\ref{eq:nz0}).

 Numerically, 
the number of weakly bound positrons can be estimated from (\ref{Q}) as follows:
\be
\label{Q1}
Q\approx 1.5\cdot 10^{6}   \left(\frac{T}{  \rm eV}\right)^{\frac{5}{4}}  \left(\frac{R}{2.25 \cdot10^{-5} \rm cm}\right)^{2}.
\ee
In what follows  we   assume that, to first order, that the  finite portion of positrons  $\sim Q$   leave the system as a result of the complicated processes mentioned above, in which case    the AQN as a system   acquires a negative  electric  charge $\sim -|e|Q$  and  get   partially  ionized as a macroscopically large object of mass $M\simeq m_pB$. 
The ratio $eQ/M\sim 10^{-19} e/m_p$ characterizing  this object  is very tiny. However, the charge $Q$ itself is sufficiently large  being capable to capture (with consequent possibility of annihilation) the  positively charged protons from surrounding ionized plasma.

  The corresponding capture radius $R_{\rm cap}(T)$ can be estimated from the condition that the potential energy of the attraction (being negative) is the same order of magnitude as kinetic energy of the protons from highly ionized gas with $x_e=1$ which is characterized by external temperature $T_{\rm gas}$, i.e.
 \be
\label{capture1}
  \frac{\alpha Q(T)}{R_{\rm cap}(T)} \sim \frac{m_pv^2}{2}\sim T_{\rm gas}  ~~ \Rightarrow ~~~  R_{\rm cap}(T)\simeq 0.2~ {\rm cm} \left(\frac{T}{  \rm eV}\right)^{\frac{5}{4}} \cdot \left(\frac{  \rm eV}{T_{\rm gas}} \right), 
  \ee
where $Q$ is estimated by  (\ref{Q}),  (\ref{Q1}).  One should emphasize that $R_{\rm cap}(T)$ depends on both temperatures, the internal $T$ through the charge $Q(T)$ as given by (\ref{Q}),  (\ref{Q1}) and external gas  temperature $T_{\rm gas}$ which is essentially determined by the typical velocities of particles in plasma and  can be identified with virial temperature of particles  in galaxy. Important  point here is that $R_{\rm cap}(T)\gg R$ such that effective cross section being $\pi R_{\rm cap}^2$ is dramatically larger than geometric size $\pi R^2$ entering (\ref{eq:rad_balance}) which would be the case if gas is  represented by    neutral atoms. In our relation (\ref{capture1}) we also neglected the Debye screening effect  as   $\lambda_D\gg R_{\rm cap}$ for all relevant values of parameters, where  $\lambda_D$ is defined as usual
 \be
  \label{Debye}
  \lambda_D\approx \sqrt{\frac{T_{\rm gas}}{4\pi\alpha n }} \sim 0.7\cdot 10^3 ~{\rm cm} \left( {\frac{T_{\rm gas}}{\rm eV}}\right)^{1/2}.
  \ee

Precisely this feature of ionization of the AQN characterized by electric charge $Q(T)$ dramatically enhances the visible-DM interaction in highly ionized environment when cosmologically relevant ratio ($\sigma/M)$
from table \ref{tab:basics} becomes large. We  illustrate this enhancement with the following estimates of this ratio  for the neutral ($x_e=0$) and highly ionized ($x_e=1$)  environments:  
\be
\label{sigma}
\frac{\sigma (x_e=0)}{M_N}\sim \frac{\pi R^2}{M_N}\sim    10^{-10}{\rm \frac{cm^2}{g}}, ~~~~~~~~~~~~ \frac{\sigma (x_e=1)}{M_N}\sim \frac{\pi R_{\rm cap}^2}{M_N}\sim    {\rm \frac{cm^2}{g} } \left(\frac{T}{ 10~ \rm eV}\right)^{\frac{5}{2}} \cdot \left(\frac{  \rm eV}{T_{\rm gas}} \right)^2.
\ee
 
 We conclude this brief overview section on   previous works on the AQNs with the following remark: all the parameters of the model as reviewed above  had been applied  previously for dramatically different studies, in very different circumstances  as highlighted in concluding section \ref{conclusion}.  We do not modify any fundamental parameters of this model  by applying this new DM scenario in next section \ref{DM-coupling} to the problem on structure formation.  In particular we shall argue that some  puzzling observations such as (\ref{correlation}) could be a direct consequence of such visible- DM  interaction (\ref{sigma}), when the AQNs  indeed behave as   {\it chameleon} like particles. To be more precise, the  effective cross section (\ref{sigma}) is highly sensitive to the density of the surrounding environment $n$, its temperature $T_{\rm gas}$  and its  ionization features $x_e$.  The corresponding parameters affect  the strength of visible-DM  interaction and its key element,    $R_{\rm cap}(T)$, which itself depends  on the environment  according to (\ref{T}) and (\ref{capture1}).  
 
 \section{DM-visible matter coupling}\label{DM-coupling}
 Our basic claim of this section is that the structure formation may be dramatically affected by the DM in form of the AQNs which strongly interact with the gas of particles. The study  of the structure formation  dynamics is obviously a prerogative of  the numerical simulations, which is the main technical tool for quantitive analysis. 
The goal of this work is less ambitious as we want to demonstrate in a pure qualitative analytical way  few important characteristics (such as portion of the ionization of the gas, $x_e$) which may be dramatically affected  by the interaction of the AQNs with surrounding plasma during the structure  formation. 
 
 To demonstrate the importance of  these key novel qualitative  effects is very  instructive to compare the analytical formula of our studies with corresponding conventional expressions which are  normally used in numerical simulations. Therefore, in what follows we use analytical formulae from the textbook \cite{Padmanabhan} as the benchmark to be compared with corresponding  AQN-based expressions.  
 We shall  demonstrate that the  DM-visible matter coupling will play the dominant role 
   in some circumstances.  The corresponding effects  may dramatically    modify the conventional picture of the structure formation.
    Our analytical studies obviously do  not replace a full scale 
 numerical simulations. However, the comparison with conventional formulae  \cite{Padmanabhan}  obviously show the basic trend which may crucially  modify some elements of the standard picture on structure formation. These modifications   have in fact, many common  consequences and manifestations with previously introduced  ad hoc models  which were    coined as  Self-Interacting dark matter, Self-Annihilating dark matter, Decaying dark matter etc. 
 
 Our comment here is that the AQN model was invented  (in contrast to all ad hoc  models mentioned above) for dramatically different purposes as overviewed  in Sect. \ref{basics} with drastically different motivation, not related in anyway to the structure formation problems being the main topic of this work. Nevertheless, there are many model-independent consequences of the AQN construction which may dramatically affect the dynamics at small scales as we argue below in section \ref{AQN-induced}.

 \subsection{Galaxy Formation. The basics picture. Notations. }\label{galaxy_formation_basics}
 The conventional picture of the structure formation assumes that CCDM particles have undergone violent relaxation such that the asymptotic density distribution $\rho_{\rm DM}(r)$  can be approximated by an isothermal sphere \cite{Padmanabhan}. If effects  of baryon's cooling are ignored the dynamical evolution of the baryons will be similar to that of DM particles. However, the collapse of baryons develops shocks and the gas get reheated to a temperature $T_{\rm gas}$ at which the pressure balance can prevent further collapse.  The corresponding temperature can be estimated as follows \cite{Padmanabhan}:
 \be
 \label{T_gas}
 \frac{3\rho_{\rm gas} T_{\rm gas}}{2 m_p}\simeq \frac{\rho_{\rm gas} v^2}{2}~~~~~ \Rightarrow ~~~~~  T_{\rm gas}\simeq \frac{m_p v^2}{3}, ~~~{\rm where}~~~ \frac{v^2}{r}\simeq \frac{GM(r)}{r^2}, 
 \ee
 where for an order of magnitude estimates and for simplification we assumed that the gas is entirely consist the hydrogen, though
 the He fraction could be relatively large. The velocity $v$ entering (\ref{T_gas}) is the circular velocity   not to be confused with mean-square velocity $\sigma_v^2=1/2 v^2$. The $T_{\rm gas}$ as defined by (\ref{T_gas}) is essentially the virial temperature $T_{\rm vir}$, but we prefer to use notation  $T_{\rm gas}$ as our goal is to   study the microscopical   interaction of this gas with the AQNs in what follows.  
 
 As the temperature $T_{\rm gas}$ becomes sufficiently high the cooling processes must be taken into account. If the cooling  processes are  sufficiently effective  the collapse may proceed further to form more tightly bound object. The corresponding evolution is entirely determined by the relative values of the dynamical time scale $t_{\rm dyn}$ and the cooling time scale $t_{\rm cool}$ defined as follows  \cite{Padmanabhan}:
 \be
 \label{t}
 t_{\rm dyn}\simeq \frac{\pi}{2}\left(\frac{R^3}{2GM}\right)^{\frac{1}{2}}\simeq 5\cdot 10^{7} {\rm yr} \left(\frac{\rm GeV \cdot cm^{-3}}{\rho_{\rm total}}\right)^{\frac{1}{2}}, ~~~
 t_{\rm cool}\equiv\frac{E}{\dot{E}}\simeq \frac{3}{2}\frac{n_{\rm gas}   T_{\rm gas}}{\Lambda(T_{\rm gas})}, ~~~  \rho_{\rm total}\equiv  (\rho_{\rm gas} +\rho_{\rm DM})
 \ee
 where $n_{\rm gas}$ is the number density of the material such that $\rho_{\rm gas} \equiv m_p n_{\rm gas}$, while $\Lambda(T_{\rm gas})$ is the cooling rate which has the meaning of energy emitted from unit volume per unit time with  dimensionality $\rm [erg \cdot cm^{-3}\cdot s^{-1}]$. In all these  estimates, including (\ref{t})     we simplify  things  and   ignore  the He portion of the gas as we already mentioned above. 
 
 The cooling rate is dominated by two   processes:  the energy loss due to the bremsstrahlung \cite{Padmanabhan}
 \be
 \label{bremsstrahlung}
 \epsilon_{\rm brem}\simeq 1.4\cdot 10^{-27}{\rm   \frac{erg}{ cm^{3}\cdot s }}\left(\frac{T_{\rm gas}}{1K}\right)^{1/2} \left(\frac{n_{\rm gas} }{\rm cm^{-3}}\right)^2   x_e^2, 
  \ee
 and the loss due to the atomic collisions \cite{Padmanabhan}
 \be
\label{collisions}
\epsilon_{\rm coll}\simeq  7.5\cdot 10^{-19}{\rm   \frac{erg}{ cm^{3}\cdot s }}  \left(\frac{n_{\rm gas} }{\rm cm^{-3}}\right)^2   x_e(1-x_e)\exp\left(-\frac{E_0}{T_{\rm gas}}\right),
 \ee
  where $x_e$ is the ionization fraction and $E_0=13.6 ~\rm eV$.  Both rates $ \epsilon_{\rm brem}$ and $\epsilon_{\rm coll}$ dramatically depend on ionization portion of the gas, the $x_e(T_{\rm gas})$ which   itself is function of the gas temperature $T_{\rm   gas}$. 
  
  The corresponding parameter $x_e(T_{\rm gas})$  is determined by the  relative strength of two competing processes: the collisional ionization which is characterized by the time scale $t_i$ and the recombination with time scale $t_r$. The corresponding time scales can be estimated as follows  \cite{Padmanabhan}:
  \be
  \label{t_i}
  t_i&\simeq& \frac{2.2\cdot 10^{14} {\rm s}}{(1-x_e)} \left(\frac{10^5K}{T_{\rm   gas}}\right)^{1/2} \exp\left(\frac{E_0}{T_{\rm gas}}\right),\\
  t_r&\simeq& \frac{5.8\cdot 10^{19} {\rm s}}{(x_e)} \left(\frac{T_{\rm   gas}}{10^5K}\right)^{2/3}.
    \label{t_r}
  \ee
Equating  (\ref{t_i}) and (\ref{t_r}) determines the ionization fraction $x_e(T_{\rm gas})$ as  a function of $T_{\rm gas}$,
\be
\label{x_e}
x_e (T_{\rm gas})\simeq \left[1+2.8\cdot 10^{-6}\left(\frac{10^5K}{T_{\rm   gas}}\right)^{7/6}\exp\left(\frac{E_0}{T_{\rm gas}}\right)\right]^{-1}.
\ee
   This expression for $ x_e (T_{\rm gas})$ can be substituted to formulae for $\epsilon_{\rm brem}$ and $\epsilon_{\rm coll}$ as given by (\ref{bremsstrahlung}) and  (\ref{collisions}) correspondingly to estimate the key parameter, the cooling rate  $\Lambda(T_{\rm gas})$ entering formula (\ref{t}). Assuming that $x_e (T_{\rm gas})\simeq 1$ one can expand (\ref{x_e}) to arrive to the following simplified  expression for the cooling rate  which is not valid for low temperatures when $x_e$ strongly deviates from unity and expansion is not justified,
   \be
   \label{Lambda}
   \Lambda (T_{\rm gas}) \simeq  10^{-24}{\rm   \frac{erg}{ cm^{3}\cdot s }} \left[0.44
   \left(\frac{T_{\rm   gas}}{10^5K}\right)^{1/2} + 2.1\left(\frac{T_{\rm   gas}}{10^5K}\right)^{-7/6} \right], ~~~ (\epsilon_{\rm coll}+  \epsilon_{\rm brem})\equiv \left(\frac{n_{\rm gas} }{\rm cm^{-3}}\right)^2  \Lambda (T_{\rm gas}).
   \ee
  In this formula   the first term in the brackets is due to bremsstrahlung  
  radiation  $\epsilon_{\rm brem}$ while the second term is the result of   atomic collisions $\epsilon_{\rm coll}$. 
   
   The cooling time scale $t_{\rm cool}$ can be estimated from (\ref{t}) as follows 
   \be
   \label{t_cool}
    t_{\rm cool}\equiv\frac{E}{\dot{E}}\simeq 2.5\cdot 10^6 {\rm yr}   \left(\frac{\rm cm^{-3}}{n_{\rm gas} }\right) \left[ \left(\frac{T_{\rm   gas}}{10^5K}\right)^{-1/2}+ 4.8  \left(\frac{T_{\rm   gas}}{10^5K}\right)^{-13/6}\right]^{-1},
   \ee
   where we literally use expression for $\Lambda (T_{\rm gas})$ as given by (\ref{Lambda}). The numerical value in the brackets for the second term (which describes the cooling due to the atomic collisions $\epsilon_{\rm coll}$) slightly deviates from the corresponding formula from the textbook  \cite{Padmanabhan}. This is  because the   expression  in \cite{Padmanabhan} was modified  to better fit with the numerical simulation results.  We opted to keep the original expressions  (\ref{Lambda}) as our main goal is the comparison of the AQN-induced mechanism with conventional mechanism  to pinpoint the   dramatic qualitative deviations from the standard cooling processes as given  by (\ref{t_cool}).

   The estimate for the cooling time scale  $ t_{\rm cool}$ allows us to compare it with the dynamical time scale $t_{\rm dyn}$ as given by (\ref{t}), where matter density  should be understood as  the {\it total} matter density of the material, including the DM portion.  
  It is convenient to define the  ratio $\cal R$ as follows:
   \be \label{R}
   {\cal R}\equiv \frac{t_{\rm cool}(\rho_{\rm gas})}{t_{\rm dyn}(\rho_{\rm total})} \leq 1, ~~~~~~ \rho_{\rm total}\equiv  (\rho_{\rm gas} +\rho_{\rm DM}),
   \ee
   when $t_{\rm cool}(\rho_{\rm gas})$ depends exclusively from hadronic gas component $\rho_{\rm gas}$ in CCDM  treatment, while 
  $ t_{\rm dyn}(\rho_{\rm total})$ depends on the total density of the material, including the DM component. 
   
 If $\cal R$  is smaller than unity the cloud will cool rapidly, and gas will undergo (almost) a free fall collapse. Fragmentation into smaller units occurs because smaller mass scales will become gravitationally unstable.   The key parameter which determines the parametrical space  when the clouds will continue to collapse defines  the region when the galaxies (and stars) may form.  Precisely this parameter  $\cal R$ governs the evolution of the system. One can estimate the typical masses, typical sizes, and the typical temperatures $T_{\rm   gas}$  of the clouds  for this collapse to happen based on analysis of domain where ${\cal R}\leq 1$.  The corresponding studies are  obviously a prerogative of the numerical N-body simulations, far outside of the scope of the present work, though  some qualitative estimates can be also made \cite{Padmanabhan}. 
 
 For the  purposes of the present work   we  use the condition (\ref{R}) as the boundary in parametrical space which specifies the domain when the galaxies may form. We will specifically pinpoint some key  modifications of this parameter $ {\cal R}$ when the AQNs are present in the system and  dramatically modify this domain. Our crucial  observation is that  $t_{\rm cool}$ will depend on both:  the visible  $\rho_{\rm gas}$ and  dark matter $\rho_{\rm DM}$ components, which qualitatively modifies condition (\ref{R}).  The corresponding  analysis   represents the topic of the next section \ref{AQN-induced}.

   One should  note  that all parameters and formulae we use  in this section as   benchmarks are taken from the textbook  \cite{Padmanabhan}.  They are obviously  outdated in comparison with the  standard parameters being used in more recent papers. Nevertheless, we opted to  use the parameters and numerics literally from  \cite{Padmanabhan} to pinpoint the key differences (in comparison with the conventional treatment of the problem) which occur when the AQN-induced interaction is taken into account. It is very instructive to understand a precise way how the new physics enters and   modifies  the standard picture in a qualitative parametrical way.  
 
  \subsection{Galaxy Formation. The AQN-induced modifications. }\label{AQN-induced}
  As we discussed in previous section the cooling rate is very sensitive to the ionization fraction $x_e(T_{\rm gas})$ as given by 
  (\ref{bremsstrahlung}) and (\ref{collisions}). In conventional picture this factor $x_e(T_{\rm gas})$ is determined by  two competing processes as represented by expression (\ref{x_e}). The main goal of this section is to demonstrate that this estimate for  $x_e(T_{\rm gas})$ is dramatically modified in the presence of the AQNs. As a result the expression for the cooling rate $ t_{\rm cool}$ as given by (\ref{t_cool}) is also changed.  These changes lead to  drastic modification of  the domain  governed by parameter ${\cal R}\leq 1 $ when  the structure formation can be formed.  One of the  main qualitative consequences of these modifications is  emergence of the relations such as (\ref{correlation}) reflecting a strong  visible-DM coupling,  which was one of the motivations for the present studies.

  The main physics process 
  which leads to such dramatic   variations can be explained as follows.  The  AQNs start to interact with surrounding material, mostly protons when the density of the gas becomes sufficiently high, around $(1~\rm cm^{-3})$. The corresponding interaction is strongly enhanced in the ionized plasma due to the long-ranged Coulomb attraction between (negatively charged) AQNs and  protons.  This enhancement of the visible- DM interaction  may dramatically {\it decrease} the ionization fraction $x_e$ of plasma due to the capturing (with consequent annihilation) of the protons from plasma and subsequent emission of   positrons from AQN's electrosphere. These positrons will  eventually  annihilate with free electrons from the plasma.  The corresponding time scale $t_{\rm AQN}$ for this process could be dramatically smaller than the recombination time scale $t_r$ estimated by (\ref{t_r}). As a result  the AQN-proton annihilation  becomes the dominant processes which dramatically reduces the ionization fraction $x_e$  of  plasma. As explained above the ionization fraction  directly affects   the domain of the parametrical space  where ${\cal R}\leq 1 $  which describes the region where  the galaxies may form.  
   
   Now we proceed with corresponding estimates. First, we estimate the number of protons being captured by AQNs per unit volume per unit time,
   \be
   \label{capture2}
   \frac{d n}{dt dV}\approx \pi R^2_{\rm cap}(T)\cdot n_p \cdot n_{\rm AQN} \cdot v_{\rm AQN},
   \ee
 where $R_{\rm cap}(T)$ is the capture radius determined by condition (\ref{capture1}) when the protons from the surrounding plasma are captured by the negatively charged AQNs and will be eventually annihilated inside the nugget's core. The formula (\ref{capture2}) explicitly shows the key role of the ionized media as the cross section for annihilation of the neutral atoms is dramatically smaller as $R\ll R_{\rm cap}(T)$. By dividing expression (\ref{capture2}) by gas number density $n$ we arrive to estimation for the frequency of the capturing (with consequent annihilation)  of the protons.  It is more convenient to represent this as time scale $t_{\rm AQN}$ for capturing of a proton from plasma by AQNs:
  \be
   \label{t_AQN}
   t_{\rm AQN}\approx \left[\pi R^2_{\rm cap}(T)  \cdot n_{\rm AQN} \cdot v_{\rm AQN}\cdot x_e\right]^{-1},~~~ n_p\equiv x_e n_{\rm gas} , ~~~ n_{\rm AQN}(r)\approx \frac{\rho_{\rm DM}(r)}{m_p \la B\ra}, 
   \ee
  where $v_{\rm AQN}\approx v$ is the same order of magnitude as the circular velocity as both are determined by pure gravitational forces (\ref{T_gas}). The time scale $ t_{\rm AQN}$ plays the key role in our discussions which follow because it competes with the recombination time scale (\ref{t_r}) when both processes decrease the ionization fraction $x_e$. 
  
  Numerically the time scale $ t_{\rm AQN}$ can be estimated as follows:
   \be
   \label{t_AQN_numerics}
   t_{\rm AQN}\approx   \frac{0.5\cdot 10^{19} {\rm s}}{(x_e)} \left(\frac{T_{\rm   gas}}{10^5K}\right)^{3/2} \cdot \left(\frac{T}{10^5K}\right)^{-5/2}\cdot\left(\frac{\rho_{\rm DM}(r)}{\rm GeV \cdot cm^{-3}}\right)^{-1},
   \ee
 which is almost one order of magnitude faster than the recombination time scale $t_r$ as estimated by (\ref{t_r}) for typical parameters ${T_{\rm   gas}}\sim {10^5K}$.
 One should comment here that the external temperature of the plasma ${T_{\rm   gas}}$ and internal temperature $T$ of the AQN's electrosphere 
 are different parameters of the system, though in the galactic environment they may assume similar numerical values. One should also note here that the recombination process and the capturing of the proton by AQNs with its consequent annihilation   inside the nugget work in the same direction by decreasing\footnote{\label{annihilation}The non-relativistic  positrons  emitted from AQNs with typical energies of order $\sim T$ will be quickly annihilated  by surrounding electrons from plasma as the cross section of the annihilation of slow electrons and positrons is of  order $\pi a_B^2\sim 10^{-16} {\rm cm^2}$, and annihilation occurs on the scales much shorter than kpc. The photons emitted due to these annihilation processes will leave the system as the corresponding cross section is very small $\sim \pi r_e^2\sim 10^{-24} {\rm cm^2}$ where $r_e\equiv \alpha/m_e$ is the electron classical radius, and the corresponding mean free path is much longer than kpc scale.} the ionization fraction $x_e$. 
 
 To estimate  the ionization fraction $x_e (T_{\rm   gas})$ when the AQN annihilation processes are operational one should equalize 
 \be
 \label{equalize} t_i^{-1}=t_r^{-1}+t_{\rm AQN}^{-1}
 \ee
 which replaces the condition $t_i=t_r$ leading to previous expression (\ref{x_e}) for  $x_e (T_{\rm   gas})$. To simplify the problem we consider the region when $t_{\rm AQN}\ll t_r$ and the recombination process can be ignored. The corresponding condition is:
  \be
 \label{t_AQN} 
 t_{\rm AQN} \ll t_r ~~~ \Rightarrow  ~~~ \left(\frac{\rho_{\rm DM}(r)}{\rm GeV \cdot cm^{-3}}\right) \cdot  \left(\frac{T_{\rm   gas}}{10^5K}\right)^{-\frac{5}{6}}  \cdot \left(\frac{T}{10^5K}\right)^{\frac{5}{2}}\gg 0.1,
  \ee
which implies that for sufficiently high DM density $\rho_{\rm DM}\sim  5 m_p n (r)$ with $n (r)\sim {\rm cm^{-3}}$  or/and sufficiently low gas temperature $T_{\rm   gas} \leq  {10^6K}$ the AQN-induced processes dominate, in which case  the ionization fraction $x_e (T_{\rm   gas})$  is entirely determined by the competition of the collisional ionization time scale $t_i$ and $t_{\rm AQN}$.

Equating $t_{\rm AQN}$  and  $t_i$  we arrive to the following estimate  for $x_e (T_{\rm   gas})$:
\be
\label{x_e_AQN-1}
x_e (T_{\rm gas})\simeq \left[1+4.4\cdot 10^{-5}\left(\frac{10^5K}{T_{\rm   gas}}\right)^{2}\cdot  \left(\frac{T}{10^5K}\right)^{\frac{5}{2}}\cdot \left(\frac{\rho_{\rm DM}(r)}{\rm GeV \cdot cm^{-3}}\right)\cdot \exp\left(\frac{E_0}{T_{\rm gas}}\right)\right]^{-1} ~~ {\rm (AQN-induced).}
\ee
This expression replaces the previous formula for $x_e (T_{\rm gas})$ in conventional scenario  (\ref{x_e}). 
Assuming that the condition (\ref{t_AQN}) is satisfied and $x_e (T_{\rm gas})\approx 1$ one can expand (\ref{x_e_AQN-1})
to arrive to the following expression for $(1-x_e)$ entering the cooling rate due to the atomic collisions (\ref{collisions}):
 \be
\label{x_e_AQN}
[1-x_e (T_{\rm gas})]\simeq  4.4\cdot 10^{-5}\left(\frac{10^5K}{T_{\rm   gas}}\right)^{2}\cdot  \left(\frac{T}{10^5K}\right)^{\frac{5}{2}}\cdot \left(\frac{\rho_{\rm DM}(r)}{\rm GeV \cdot cm^{-3}}\right)\cdot \exp\left(\frac{E_0}{T_{\rm gas}}\right).    
\ee
 
 Now we can substitute this expression for $(1-x_e)$ to the formula  (\ref{collisions}) for the cooling rate due to the atomic collisions:
 \be
\label{collisions-AQN}
\epsilon_{\rm coll}  \simeq 3.4\cdot 10^{-23}{\rm   \frac{erg}{ cm^{3}\cdot s }} \left(\frac{n_{\rm gas} }{\rm cm^{-3}}\right)^2 
 \cdot\left(\frac{T_{\rm   gas}}{10^5K}\right)^{-2}\cdot  \left(\frac{T}{10^5K}\right)^{\frac{5}{2}}\cdot \left(\frac{\rho_{\rm DM}(r)}{\rm GeV \cdot cm^{-3}}\right)~~ {\rm (AQN-induced).}
\ee
 At the same time the expression for   cooling due to the bremsstrahlung radiation $ \epsilon_{\rm brem}$ remains the same as it is not sensitive to $x_e$ as long as it is close to unity. As a result, the expression for the cooling rate  $\Lambda (T_{\rm gas})$ assumes the form
 \be
 \label{Lambda_AQN}
  \Lambda (T_{\rm gas}) \simeq  10^{-24}{\rm   \frac{erg}{ cm^{3}\cdot s }} \left[0.44
   \left(\frac{T_{\rm   gas}}{10^5K}\right)^{\frac{1}{2}} +  34 \left(\frac{T_{\rm   gas}}{10^5K}\right)^{-2}   \left(\frac{T}{10^5K}\right)^{\frac{5}{2}} \left(\frac{\rho_{\rm DM}(r)}{\rm GeV cm^{-3}}\right) \right] , 
 \ee
 where the first term due to the  bremsstrahlung radiation  remains the same as in the conventional treatment (\ref{Lambda}), while the second term describing  the atomic collisions is dramatically larger by one order of magnitude than in (\ref{Lambda}). The basic reason for this difference is that 
 the cooling rate due to the atomic collisions is proportional to density of the neutral atoms $n_H\propto  (1-x_e)$ which increases in the presence of the AQNs  in comparison with conventional case due to the mechanism described at the very beginning of this section.

 The increase  of the cooling rate $\Lambda (T_{\rm gas})$ leads to consequent dramatic modification in the cooling time scale $t_{\rm cool}$ as defined by  (\ref{t_cool}). In the presence of the AQNs in the system it gets modified as follows:
  \be
   \label{t_cool_AQN}
    t_{\rm cool}\equiv\frac{E}{\dot{E}}\simeq 2.5\cdot 10^6 {\rm yr}   \left(\frac{\rm cm^{-3}}{n_{\rm gas} }\right) \left[ \left(\frac{T_{\rm   gas}}{10^5K}\right)^{-1/2}+ 77 \left(\frac{T_{\rm   gas}}{10^5K}\right)^{-3}   \left(\frac{T}{10^5K}\right)^{\frac{5}{2}} \left(\frac{\rho_{\rm DM}(r)}{\rm GeV \cdot cm^{-3}}\right) \right]^{-1},
   \ee
   where the first term in the brackets due to the  bremsstrahlung radiation  remains the same as in the conventional treatment  (\ref{t_cool}), while the second term describing  the atomic collisions is dramatically enhanced in comparison with (\ref{t_cool}) due to the same reasons described above.  
   
   There are two key points here: first, $ t_{\rm cool}$ depends on $\rho_{\rm DM}(r)$ which is a highly nontrivial new qualitative effect  
   because in conventional  CCDM picture any  cooling effects fundamentally cannot depend on $\rho_{\rm DM}(r)$ as these effects   are entirely determined by the visible baryonic matter in form of the gas. The second point here 
    is that for the temperatures $T_{\rm   gas}\leq 10^{5} K$ the cooling  time scale $t_{\rm cool}$  is one order of magnitude shorter than in  conventional treatment (\ref{t_cool}). In fact, the AQN-induced processes remain to be the dominant cooling mechanism up to $T_{\rm   gas}\simeq 10^6K$ for $\rho_{\rm DM}\approx 5 \rho_{\rm gas}$, and could remain the dominant mechanism even    for higher temperatures. 
 
 The dramatic modifications in $t_{\rm cool}$ implies that the key parameter  ${\cal R}$ will also experience crucial qualitative changes in defining of   the domain where the structure formation may form,   
   \be 
   \label{R_AQN}
   {\cal R}\equiv \frac{t_{\rm cool}(\rho_{\rm gas}, \rho_{DM})}{t_{\rm dyn}(\rho_{\rm total})} \leq 1, ~~~~~~ \rho_{\rm total}\equiv  (\rho_{\rm gas} +\rho_{\rm DM}).
   \ee
Indeed,  in contrast  with the original definition (\ref{R}) the cooling time   $t_{\rm cool}(\rho_{\rm gas}, \rho_{DM})$ entering (\ref{R_AQN}) now depends explicitly on both,  the $\rho_{\rm gas}$ and $ \rho_{DM}$ as expression (\ref{t_cool_AQN}) explicitly shows. 
 
 We conclude this section  with the following generic comment. We have made a large number of technical assumptions in this section to simplify the analysis to argue that  the visible-DM interaction may dramatically modify some parameters, such as $x_e$ and cooling time $ t_{\rm cool}$  as a result of the AQN-induced processes. It was not the goal of this work to perform a full scale simulations and modelling, which is well beyond the scope of the present work. However, we observed a number of qualitative features of the system, which are known to occur,  but   cannot be easily understood  within conventional models of structure formation. In next section \ref{observations} we briefly overview  some  observational consequences of the AQN framework, which are hard to  understand within conventional models, but could be easily understood 
 within a new paradigm when the baryonic  and DM constituents    become strongly interacting components of the system.

 \section{Observable consequences of the visible-DM interaction. The New Paradigm.}\label{observations}
 As we reviewed in Sect.\ref{AQN} the AQN model is dramatically distinct  from conventional DM proposals as the central elements of the DM configurations are the same (anti)quarks and gluons from QCD which represents the inherent part of  the Standard Model (SM).   Therefore,  the AQNs  become the  strongly interacting objects in sufficiently dense environment. In other words, the AQN behaves as a {\it chameleon}: it does not interact with the surrounding material in dilute neutral environment, but it becomes strongly interacting object in sufficiently dense environment. Therefore, the AQN framework  essentially represents an explicit realization  of the New Paradigm, when the visible and  DM  building blocks become strongly interacting  components if some conditions are met. It must be contrasted with    conventional CCDM paradigm when these distinct components, by definition,  never couple. 
 
 The main purpose  of this section is to briefly overview  a number of qualitative consequences of this new paradigm.  
  The corresponding properties  which are listed below are not very sensitive to any specific numerical values  
 of the parameters used in the previous sections. Instead, these novel features represent the inherent features of the new framework. 
 In next subsection \ref{sect:correlation} we explain in qualitative way how the observed correlation such as  (\ref{correlation})
 could emerge in the New Paradigm. In section \ref{sect:Heater} we argue that the new paradigm inevitably implies 
 emission of the  additional energy in different  frequency bands, including  the UV  radiation. Apparently, there are several recent hints from JWST, see e.g. \cite{JWST,ALMA,Labbe,Boylan-Kolchin:2022kae}  that such excess of radiation has been indeed observed 
at large $z$.    In section \ref{sect:Heater1} we argue that the  very same  effects   could be responsible for   the mysterious diffuse UV radiation    \cite{Henry_2014,Akshaya_2018,2019MNRAS.489.1120A}  at present time $z=0$ as suggested in  \cite{Zhitnitsky:2021wjb}. 
 
 \subsection{How the observed correlation (\ref{correlation}) could emerge in the New Paradigm?}\label{sect:correlation}
 To simplify our qualitative analysis we consider the domain in the parametrical space when the AQN-induced term dominates  the  conventional cooling due to the bremsstrahlung radiation. The corresponding condition can be estimated from (\ref{t_cool_AQN}) as follows:
 \be
 \label{AQN-domination}
 \left(\frac{T_{\rm   gas}}{10^6K}\right)^{-\frac{5}{2}}   \left(\frac{T}{10^5K}\right)^{\frac{5}{2}} \left(\frac{\rho_{\rm DM}(r)}{\rm GeV \cdot cm^{-3}}\right)\gg 4.1,  
 \ee
 which is satisfied in the region with $T_{\rm   gas}\leq {10^6K}$ and $\rho_{\rm DM}\geq  5 \rho_{\rm gas}$ with our benchmark density $ n_{\rm gas}\approx \rm cm^{-3}$. This parametrical region is largely   overlap with condition (\ref{t_AQN})  that the AQN-induced processes dominate 
 conventional recombination effects in computations of the ionization fraction $x_e$ such that all our simplification and estimates remain consistent.  
 
 Therefore, assuming the condition (\ref{AQN-domination}) is satisfied our basic requirement (\ref{R_AQN}) defining the domain $ {\cal R}\leq 1$ where galaxy may form can be written in the following simple way 
 \be
 \label{AQN-correlation}
  \frac{[\tilde{\rho}_{\rm DM} \cdot \tilde{\rho}_{\rm gas} ]}{ \sqrt{ \tilde{\rho}_{\rm DM} + \tilde{\rho}_{\rm gas} }}    \left(\frac{T_{\rm   gas}}{10^6K}\right)^{-3}   \left(\frac{T}{10^5K}\right)^{\frac{5}{2}}\geq 6.5, ~~~~ \tilde{\rho}_{\rm DM} \equiv \left(\frac{\rho_{\rm DM}(r)}{\rm GeV \cdot cm^{-3}}\right), ~~~~   \tilde{\rho}_{\rm gas} \equiv \left(\frac{\rho_{\rm gas}(r)}{\rm GeV \cdot cm^{-3}}\right),
 \ee
 where we use formula (\ref{t}) for the $t_{\rm dyn}$ and expression (\ref{t_cool_AQN}) for $t_{\rm cool}$ assuming the AQN-induced term dominates according to the condition (\ref{AQN-domination}).
 
 The crucial point here is that the visible-DM interaction representing the key element of a new paradigm explicitly manifests  itself  
 in formula (\ref{AQN-correlation}) which strongly resembles the structure of the correlation (\ref{correlation}) inferred from the observations long ago, see review \cite{Salucci:2020eqo}. The coupling of the visible and dark components is explicitly present in the system, and formula (\ref{AQN-correlation}) is a direct consequence of this interaction. 
 
  It is important to emphasize that the condition (\ref{AQN-correlation})  is local in nature as it depends on $[\tilde{\rho}_{\rm DM}(r) \cdot \tilde{\rho}_{\rm gas}(r) ]$ in region $r\leq r_0$ where the densities are sufficiently large, roughly $\rho_{\rm DM}(r)\sim \rho_{\rm gas}(r)\sim 10^{-24}   {\rm (g \cdot cm^{-3}})$ in cgs units. For $r\gg r_0$ the the visible-DM interaction can be ignored and the AQNs behave in all respects as CCDM. The locality of the condition (\ref{AQN-correlation}) implies that region $r_0$ where the visible-DM interaction becomes the dominant element of the system does not depend on the size of the cloud which is about to collapse to form a galaxy, nor on its mass, which is precisely the claim of \cite{Salucci:2020eqo} as stated in (\ref{correlation}). 
  
  Parameter $r_0$ was identified with the size of the core in \cite{Salucci:2020eqo}, while in our microscopical treatment of the system the scale $r_0$ is identified with condition (\ref{AQN-correlation}). Therefore, the core formation  in the AQN framework can be interpreted as  a result of strong visible- DM interaction when condition (\ref{AQN-correlation}) starts to  satisfy at $r\approx r_0$. 
  
  The temperature $T_{\rm   gas}$ entering (\ref{AQN-correlation}) can be thought as the virial temperature as defined by (\ref{T_gas}), which indeed assume the values in the range $T_{\rm   gas}\simeq (10^5-10^6)K$. Another parameter $T$ entering (\ref{AQN-correlation})  is the internal temperature of the AQNs, and should not be confused 
  with $T_{\rm   gas}$. This parameter is very hard to estimate as reviewed in section  \ref{AQN}, but it must also lie in the range $T\simeq (10^5-10^6)K$ for the environment under consideration. We shall not elaborate in details on this matter in the present work.   
 
 \subsection{The AQNs as the  UV emitters  in early  galaxies}\label{sect:Heater}
 The processes which lead to the correlation  (\ref{AQN-correlation}) as discussed above are always accompanied by the radiation in many different frequency bands as we discuss below. Indeed, the total amount of energy being produced per unit time is determined by the right hand side of   (\ref{eq:rad_balance}). This energy will be released into the space in many different forms, including the axion and neutrino emissions.
 However, the dominant portion of the emission will be in form of radiation from electrosphere according to (\ref{eq:P_t}). 
 The spectrum of the radiation is very broad $\omega \leq T$ as computed in  \cite{Forbes:2008uf} and depicted on Fig.  \ref{AQN-structure}.  
 This is the dominant radiation process. There is also annihilation of the emitted positrons with electrons from plasma as mentioned in footnote \ref{annihilation}. However, the total released energy due to these annihilation processes is obviously suppressed by factor $m_e/m_p\ll 1$. In what follows we estimate the total amount of energy being produced as a result of the annihilation processes during the galaxy formation.
 
 We start from expression (\ref{capture2}) for number of annihilation events per unit time per unit volume. We multiply this expression by factor 
 $2m_pc^2\simeq 2 \rm ~GeV$ to arrive to estimate for energy being produced as a result of these annihilation processes: 
  \be
   \label{energy-released}
   \frac{d E}{dt dV}\approx \pi R^2_{\rm cap}(T)\cdot n \cdot n_{\rm AQN} \cdot v_{\rm AQN} \cdot x_e\cdot ( \rm  2~ GeV)
   \ee
 By dividing expression (\ref{energy-released}) by gas number density $n$ we arrive to estimation of the energy being released per unit time    per  single  proton (or hydrogen atom) in plasma. To estimate the total energy being released by this mechanism we have to multiply the estimate (\ref{energy-released}) by the Hubble time at redshift $z$. Thus, we arrive to an order of magnitude estimate for the total energy released by the AQNs due to the annihilation events during the Hubble time  per  single  proton (or hydrogen atom) in plasma:
   \be
   \label{energy-per-proton}
     \frac{d E}{dt} H^{-1}\sim  { 4\cdot 10^{-7}}~{\rm GeV}\cdot  x_e\cdot  \left(\frac{T_{\rm   gas}}{10^6K}\right)^{-3/2} \cdot \left(\frac{T}{10^5K}\right)^{5/2}\cdot\left(\frac{\rho_{\rm DM}(r)}{10^{-3}~\rm GeV \cdot cm^{-3}}\right),
   \ee
 which of course represents a tiny  portion ($\sim 10^{-7}$) in comparison with $m_pc^2$.  In  estimate (\ref{energy-per-proton})  we use $H^{-1}\simeq 10^9 \rm yr$. 
This estimate suggests   that the total amount of the DM as well as the typical  size of the AQNs will  not be  affected  during the Hubble time due to the annihilation processes.  

Few comments are in order. First of all, an order of magnitude estimate (\ref{energy-per-proton}) should be considered as an upper limit for the energy released due to the annihilation events. Indeed,  there  are many other forms of the emission such as the axion and neutrino emissions. Furthermore, a typical internal temperature $T$ of the AQNs   in peripheral   regions  of the galaxy    is well below than $10^5K$   as the number of annihilation events in these regions  is very tiny, which also decreases the estimate   (\ref{energy-per-proton}).  
 The peripheral   regions do not contribute to the emission at all, as the AQNs behave as conventional  CCDM particles outside of the cores, as we already mentioned. All  these effects obviously  further reduce the total energy being released due to the annihilation events. 
 
 The most important comment here is as follows. The dominant portion of the electromagnetic   emission (\ref{energy-per-proton})  is very broadband with   typical frequencies around $\omega\leq T\simeq 10^5K $. As a result,   the UV radiation is expected to occur from the AQN-induced processes as a typical internal temperature  could reach values $T\simeq 10~ \rm eV$ or even higher. This UV emission is very generic consequence of the AQN framework, and always occurs  in addition to  the UV emission from the  stars, which is considered to be conventional source of the UV radiation in galaxies. The intensity of this emission in the AQN framework should be about the same order of magnitude as the observed excess of the diffuse UV emission in our galaxy, see next subsection  \ref{sect:Heater1}.
 
 It is interesting to note that   there are several recent hints from JWST, see e.g. \cite{JWST,ALMA,Labbe,Boylan-Kolchin:2022kae}, suggesting that the  excess of the UV radiation  from red-shifted  galaxies has been indeed observed. 
  In fact, it has been argued that the UV luminosity would need to be boosted by about a factor of $\sim 2.5$ to match the observations  at $z\sim 11$ according to \cite{JWST}. If these observations will be confirmed by future analysis it could support our interpretation that the observed excess of the UV emission could be  due  to the AQN annihilation processes.
  \exclude{ 
 Our  interpretation on excess of the UV radiation in red-shifted galaxies  can be further supported by recent analysis of the diffuse UV emission 
 in present time in our galaxy where similar excess of the diffuse UV emission 
 had been also observed. We shall overview these observations and possible explanations within the AQN framework in the next subsection  
 \ref{sect:Heater1}.
 }
 \subsection{The AQNs as the  UV emitters   at present time}\label{sect:Heater1}
Our claim that the  excess of the UV emission must be present in all galaxies can be tested in our own galaxy by studying the detail characteristics of the diffuse UV emission.
In fact such studies had been recently carried out in  \cite{Henry_2014,Akshaya_2018,2019MNRAS.489.1120A}. The corresponding results 
   are very hard to understand if interpreted in terms of the conventional astrophysical  phenomena
when the dominant source of the diffused  UV background  is the dust-scattered radiation of the UV emitting stars.  The  analysis \cite{Henry_2014,Akshaya_2018,2019MNRAS.489.1120A}  very convincingly disproves   this conventional picture.  The arguments are based on a number of very puzzling observations which are not compatible with standard picture. We mention here just two of these mysterious observations: 1. The diffuse radiation is very uniform in both hemispheres, see Figs 7-10 in \cite{Henry_2014}. This feature should be contrasted to the strong non-uniformity in distribution of the UV emitting stars; 2. The diffuse radiation is almost entirely independent of Galactic longitude. This feature must be contrasted with localization of the brightest UV emitting stars which are overwhelmingly confined to the longitude range $180^0-360^0$. These and several similar observations   strongly suggest that the diffuse background radiation can hardly  originate in dust-scattered starlight.  The authors of \cite{Henry_2014} conclude that the  source of the diffuse FUV emission is unknown --that is the mystery that is referred to in the title of the paper \cite{Henry_2014}.

We proposed in  \cite{Zhitnitsky:2021wjb}  that  this excess in UV radiation   is the result of the 
    dark matter annihilation events  within  the AQN dark matter 
    model.  The   excess of the UV radiation observed at $z=0$ and studied in  \cite{Henry_2014,Akshaya_2018,2019MNRAS.489.1120A} has precisely the same nature and originated from the same source in form of the dark matter AQNs as advocated in this work for red-shifted galaxies as overviewed in previous subsection \ref{sect:Heater}. The proposal  \cite{Zhitnitsky:2021wjb}  is supported by demonstrating that   intensity and the spectral features of the   AQN induced emissions    are consistent with  the corresponding  characteristics of the observed excess  \cite{Henry_2014,Akshaya_2018,2019MNRAS.489.1120A}  of the UV radiation.

The excess of the UV radiation measured by GALEX  over its bandpass (1380-2500)$  \AA$  varies between 
$(300- 1800) \cdot [{\rm photons   ~  {cm}^{-2}~ {s}^{-1} sr^{-1}}\AA^{-1}]$ depending  on the galactic latitude, see Fig. 14 in \cite{Henry_2014}. One can represent  this measurement in conventional units as follows
\be
\label{units_conversion} 
I^{\rm FUV}_{\nu}\approx (300- 1800) \cdot \int^{2500}_{1380}\rm  d\lambda\frac {h\nu}{cm^2\cdot {s}  \cdot sr \cdot \AA }\approx 
(3.6-22)\cdot 10^{-6}~ \frac{ erg}{cm^{2}\cdot {s}  \cdot sr }\approx(3.6-22) \frac{nW}{m^2 \cdot sr},  
\ee
where the photon's count  was   multiplied by factor $h\nu = {hc}/{\lambda}$ and integrated over  its bandpass (1380-2500)$\rm \AA$ assuming  the flat spectrum. The observed  intensity (\ref{units_conversion}) is consistent with the AQN proposal \cite{Zhitnitsky:2021wjb}. We expect that a similar intensity could  contribute to the UV emission (in addition to the luminosity from the UV emitting stars) of the red-shifted galaxies    as mentioned at the very end of the previous subsection \ref{sect:Heater}.

We conclude this section with the following generic comment. We advocate a new paradigm that the visible and  DM  components become strongly interacting  components if some conditions are met. This    should be contrasted with    conventional CCDM paradigm when DM and visible matter components never interact.  The new paradigm  has many observable consequences, such as emergence of the correlations  (\ref{correlation})  
mentioned in subsection \ref{sect:correlation} and excess of the diffuse UV   emission  (along with radiation in other frequency bands) as highlighted in subsections \ref{sect:Heater} and \ref{sect:Heater1}. There are many more mysterious and puzzling observations at dramatically different scales in different systems which   also suggest that a new source of energy apparently is present in variety of  systems, from galactic scale to the Sun and  Earth, which is the topic of the Concluding comments of section \ref{conclusion}.

\section {Concluding comments and Future Developments} \label{conclusion}
 The presence of the {\it antimatter} nuggets\footnote{We remind the readers that the antimatter in this framework appears as natural resolution of the  long standing puzzle on similarity  between visible and DM components, $\Omega_{\rm DM}\sim \Omega_{\rm visible}$  irrespectively to the parameters of the model. 
 This feature is a result of the dynamics of the  $\cal{CP}$ violating axion field during the QCD formation period, see Sect. \ref{AQN} for review. No any other DM models, including the original Witten's construction   \cite{Witten:1984rs} can provide such similarity between these two  matter components of the Universe without fine-tunings.}  in the system implies, as reviewed in Sect.\ref{AQN}, that there will be  annihilation events (and continues injection of energy at different frequency bands,  from UV to the radio bands) leading to  large number of observable effects on different scales: from Early Universe to the galactic scales to the Sun and the  terrestrial rare events. 

 The structure formation dynamics, which is the topic of this work,  is obviously a prerogative of the numerical simulations. However, our goal in this work was to  pinpoint   the     elements in the dynamics of the galaxy formation where AQN-induced processes   dominate the dynamics, and   dramatically   modify  the structure  at small scales. In Sect. \ref{sect:results} below we   list the basic claims of our studies.  In Sect. \ref{sect:tests} we list some possible new tests which can substantiate or refute this new paradigm.    Finally,   in Sect. \ref{sect:paradigm} we describe  several  other mysterious and puzzling observations, which can be understood within the same AQN framework,  and which indirectly support  our proposal. The evidences mentioned in Sect. \ref{sect:paradigm} are collected   in dramatically different   environments such as the  Early Universe,  post-recombination epoch, solar corona, Earth's atmosphere.

     \subsection{Basic results. }\label{sect:results}
     Our basic results can be summarized as follows. We argued in Sect.  \ref{sect:correlation} that the observed correlation such as  (\ref{correlation})   could naturally emerge in the AQN framework.  The condition (\ref{R_AQN}) defining the domain
     when galaxies may form  (\ref{AQN-correlation})  assumes dramatically  different structure  because   the cooling time   $t_{\rm cool}(\rho_{\rm gas}, \rho_{DM})$ entering (\ref{R_AQN}) now depends explicitly on both,  the $\rho_{\rm gas}$ and $ \rho_{DM}$, which is qualitatively distinct feature from conventional picture when  $t_{\rm cool}(\rho_{\rm gas})$ could only  depend  on the visible baryonic component.  The basic technical reason for such dramatic modification of the cooling rate (in comparison with conventional estimates) is due to the decreasing of the ionization fraction $x_e$ in the presence of the dark matter AQNs at small scales. 
     
          Our hope is that many of the puzzling problems listed  in Sect.\ref{sec:introduction} (such as Core-Cusp problem etc), including  correlation (\ref{correlation}) may find their resolutions if this new feature of the system will be incorporated in future simulations on structure formation.

     The processes which lead to the  condition (\ref{AQN-correlation})  where the galaxies may form will be   always accompanied by injection of some  energy  in different frequency bands (from radio to UV)   in the same  parts of the galaxies  due to the annihilation processes  in that regions.   The estimate (\ref{energy-per-proton}) provides an upper limit   for the total released energy during the Hubble time per single proton (hydrogen atom).  In particular, one can argue that the excess of the UV emission which is  apparently observed in red-shifted galaxies and in our own galaxy could be related to these annihilation processes as mentioned in sections \ref{sect:Heater} and \ref{sect:Heater1} correspondingly. 

 \subsection{New possible tests}\label{sect:tests}
 The main   element of the AQN framework  is that all new effects
 are determined by the line of sight  $\Gamma$ which includes  both:
the DM and visible matter distributions:
\begin{equation}
  \label{eq:direct_integral}
  \Phi_{\Gamma}  \propto 
  \int_{\Gamma}   d{l}\, \rho_{\rm gas} (l)\rho_{\rm DM}(l),
\end{equation}
which is inevitable feature of the   framework when the DM consist  the AQNs being made from the same strongly interacting quarks and gluons   the baryonic matter made of. The coefficient of the proportionality (the strength of the interaction) is very sensitive to  the environment.   It becomes strong at small scales when the density of the gas is relatively large. It is a highly non-linear effect as emphasized  in overview   section \ref{AQN}.

The eq. (\ref{eq:direct_integral})  is precisely the coupling we used in all our estimates starting with (\ref{capture2}).
Exactly this interaction  leads to all  dramatic consequences at small scales mentioned above in subsection \ref{sect:results}.
 The coupling (\ref{eq:direct_integral}) should be   contrasted with conventional WIMP-like models when DM and visible components do not couple. Some modifications of the DM models, such as  Self-Interacting dark matter, Self-Annihilating dark matter,
  Decaying dark matter, and many others,  depend on  DM distribution, but not on visible component. 
   Therefore, some specific morphological correlations of the  DM and visible matter distributions originated from (\ref{eq:direct_integral}) can be explicitly studied in future. The same formula  (\ref{eq:direct_integral})  essentially determines the intensity and the spectral features  of the emission due to this visible-DM coupling. 
  
  In particular, as we mentioned in section \ref{sect:Heater1} the  well-established  excess of the  diffuse UV emission   \cite{Henry_2014,Akshaya_2018,2019MNRAS.489.1120A}  which cannot be explained by conventional astrophysical sources could be   naturally understood  within the AQN framework  \cite{Zhitnitsky:2021wjb}. In this case one can study the morphology of the DM and visible matter distributions as well as ionization features of the clouds along the line of sight $\Gamma$  \cite{FUV-simulations}. 
  
  Furthermore,  one could expect   a similar     emission to occur in red shifted galaxies (of course with  rescaling of the corresponding frequency bands for non-vanishing  $z$) as mentioned in \ref{sect:Heater}. This is  because the source  of the emission   excess in red-shifted galaxies and in our own galaxy is one and the same and it is originated from the same AQN annihilation processes.   The luminosity of the AQN-induced FUV emission was estimated in  \cite{Zhitnitsky:2021wjb} and it is consistent with observed intensity as given by (\ref{units_conversion}), while conventional WIMP-like models can generate the intensity which is 17 orders of magnitude smaller than observed \cite{Henry_2014}.
  
  \exclude{
    Another possible test of this AQN framework is to study the morphology of the optical excess diffuse emission  as observed  in \cite{Lauer_2022}.  If the  ``anomalous flux" is indeed originated from the AQN induced annihilation processes  as suggested  in Sect. \ref{sect:Heater2},
    it must be correlated with FUV excess emission within AQN framework. In particular, it must demonstrate the same   galactic latitude dependence, similar to the FUV analysis  \cite{Henry_2014}. The corresponding studies   can be carried out  utilizing  the same technical tools as used  in  \cite{FUV-simulations}. 
}

     \subsection{Other (indirect) evidences for a new Paradigm}\label{sect:paradigm}
     
  There are many hints (outside the galactic scale) suggesting that the  annihilation events (which is inevitable feature of this framework) may  indeed  took place in early Universe as well as   in present epoch. In particular, the  AQNs do not affect BBN production for H and He, but   might be responsible for a resolution of   the  ``Primordial Lithium Puzzle" due to its   large electric charge $Z=3$, see \cite{Flambaum:2018ohm} for the details. In addition to the UV excessive radiation mentioned at the very end of Sect. \ref{observations},
the AQNs  may also  help to alleviate the tension between standard model cosmology and the recent EDGES observation of a stronger than anticipated 21 cm absorption feature as argued in \cite{Lawson:2018qkc}.   The AQNs might be also responsible for famous   long standing problem of  the ``Solar Corona Mystery"
  \cite{Zhitnitsky:2017rop,Raza:2018gpb} when the   so-called ``nanoflares" conjectured by Parker long ago \cite{Parker} are   identified with the  annihilation events in the AQN framework. 
  
  The corresponding very rare AQN-induced  events on Earth cannot be studied by any conventional   DM instruments   today because their small sizes such that the corresponding AQN flux is at least 20 orders of magnitude smaller than the  WIMP's flux due to very heavy nugget's mass as reviewed in Section \ref{AQN}. However, the cosmic ray (CR) laboratories  with typical size of 100 km are capable to study such rare events. 
  
 In fact,  there are several   unusual and mysterious  observations of the    CR-like events which might be related to the AQN propagating in atmosphere.    In particular,  it includes  the   \textsc{ANITA}  observation    \cite{Gorham:2016zah,Gorham:2018ydl}  of two anomalous events    with noninverted polarity which can be explained within AQN framework \cite{Liang:2021rnv}. It also includes    the  Multi-Modal Clustering  Events observed by HORIZON 10T \cite{2017EPJWC.14514001B,Beznosko:2019cI} which impossible to understand  in terms of the CR events, but  which could be  interpreted in terms of  the  AQN annihilation events in atmosphere   as argued in  \cite{Zhitnitsky:2021qhj}. Similar mysterious  
 CR-like events had been also recorded by AUGER \cite{Zhitnitsky:2022swb} and Telescope Array \cite{Zhitnitsky:2020shd}. 
The  CR-like  events can also manifest themselves in form of the acoustic and seismic signals  \cite{Budker:2020mqk}, and could be in principle recorded if dedicated instruments are present  in the same area where CR detectors are located. In this case the synchronization between different types of instruments could play a vital role in the discovery of the DM.

  The same DM source in form of the AQNs  could resolve a number of different, but related puzzles. In particular, the same AQNs could be the source of ionizing radiation which is  known to be present  well above the galactic plane \cite{Henry:2018yar}. Furthermore, the same DM source in form of the AQNs may also contribute to the resolution of  another long standing problem 
related to  the Extragalactic Background Light (EBL). Indeed, it has been known for some time that the conventional measurements    cannot be explained by diffuse galaxy halos or intergalactic stars. The discrepancy could be as large as  factor $\sim (2-3)$ or even more, see e.g. recent reviews \cite{COB,Mattila:2019ybk}. Our comment here is that the  
AQNs may fulfill this shortage as the energy injection at different scales is an inevitable feature of this construction, as the {\it antimatter} nuggets (along with matter nuggets) represent the  DM density in this framework, see Sect \ref{AQN}.  Furthermore,   the spectrum   of the emission  is very broad and  includes UV, optical, IR light, and even the radio frequency bands.  

On the observational side,  there are indeed a numerous number of hints  suggesting the   excessive   diffuse radiation  in many  frequency bands, from UV to the radio emissions.   As the latest examples one could mention  the observed    ``anomalous flux"   \cite{Lauer_2022} of the COB excess.  One could also mention that the observed widths of Ly-$\alpha$ forest absorption lines are much wider compared to conventional numerical simulations. Observations suggest that there is a non-canonical heating process in IGM neglected in simulations such that an additional energy $\sim 6.9 \rm ~eV$ per baryon is required to match the observations, which can be done e.g. with dark photon DM model \cite{PhysRevLett.129.211102}. Another recent example is the observed excess in radio frequency bands, $\nu\in \rm (150 ~MHz, 8.4 ~GHz)$ where significant discrepancy remains as large as factor $\sim 5$ \cite{Tompkins_2023}.

   Our comment here is as follows. Every single mysterious  and puzzling effect  mentioned above can be in principle explained with some specifically designed model with specifically chosen parameters such as \cite{PhysRevLett.129.211102}. 
   In contrast,  the AQN model was invented long ago     \cite{Zhitnitsky:2002qa} with dramatically different motivation for drastically different purposes.    
    Nevertheless, a large number of  puzzling, mysterious and anomalous events mentioned above could be simultaneously explained within the same framework with the same set of parameters.     In particular, the required energy injection  to explain the observed widths of Ly-$\alpha$ forest   could be naturally  explained by the AQN annihilation processes with very broad spectrum and total energy estimated by (\ref{energy-per-proton}). The same energy injection could be responsible for EBL excesses in different frequency bands, as mentioned above.
       
   We conclude this work with the following final comment.
   We advocate an idea that  the basic paradigm on nature of DM should be changed: from old paradigm (when DM is  non-baryonic weakly interacting particles)  to new paradigm when DM is baryonic and strongly interacting composite  system, made from (anti)quarks and gluons 
   of the Standard Model as reviewed in Sect.\ref{AQN}.  The AQNs in this framework behave as {\it chameleon}-like particles: they behave as conventional DM components in low  density environment, and become strongly interacting  macroscopically large objects in relatively  high density environment. The new paradigm  has many   consequences which are mentioned  above and in Sect.\ref{observations}, and which  are consistent with all presently available cosmological, astrophysical, satellite and ground-based observations.   In fact, it may even shed some light  on the  long standing puzzles and mysteries as mentioned above in Sect.\ref{sect:paradigm}.

   The structure formation dynamics, which is the topic of this work,  is obviously a prerogative of the numerical simulations  as we mentioned many times in the text. The goal of the present work with analytical estimates is to pinpoint the  exact places and elements in the dynamics of the galaxy formation where AQN-induced processes become dominating and lead to a dramatic deviation at small scales  from the conventional paradigm. Therefore, with this work we advocate  the community to incorporate this new crucial  element  on visible -DM interaction (\ref{eq:direct_integral}) in   the numerical simulations.
   
     If future  observations along with numerical simulations (which would incorporate   the  visible -DM interaction) will   confirm and substantiate  the basic consequences of this work as  listed above and in Sect.\ref{observations}    it would represent  a strong     argument  suggesting that  the resolution of   two long standing puzzles in cosmology -- the nature of the DM   and the   matter-antimatter asymmetry of  our Universe--  are  intimately  linked.    The corresponding deep connection is  automatically implemented  and incorporated in the AQN framework by its construction
     as briefly overviewed in Sect.\ref{AQN}.   

     \section*{Acknowledgements}
    I am thankful to Joel Primack for long and very useful discussions on many different aspects of the new paradigm advocated in the present work. I am also thankful to Ludo Van Waerbeke for collaboration on many completed and ongoing projects related to this new paradigm  and for  very useful explanation on how the astro/cosmology community  (astro-ph in terms of the arXiv nomenclature)  operates, which  is very different from  hep-ph   physics community practices.  This research was supported in part by the Natural Sciences and Engineering
Research Council of Canada.

\appendix

\section*{References}

  \bibliography{galaxy-formation}

\end{document}